\documentclass[a4paper,conference]{IEEEtran}
\IEEEoverridecommandlockouts
\usepackage{stfloats,color}
\usepackage{amsfonts}
\usepackage{amsmath}
\usepackage{dsfont}
\usepackage{setspace}
\usepackage{adjustbox}
\usepackage{ragged2e}
\usepackage{cite}
\usepackage{bbm}

\usepackage{filecontents}
\usepackage{array}

\usepackage{mdwmath}
\usepackage{multirow}
\usepackage{makecell}
\usepackage{times}
\usepackage{diagbox}
\usepackage{epsfig}
\usepackage{latexsym}
\usepackage{epstopdf}
\usepackage{verbatim}
\usepackage{units}
\usepackage{amsthm}
\usepackage{mdwlist}
\usepackage{esvect}
\usepackage{tabularx,colortbl}

\usepackage[unicode=true, linktocpage, linkbordercolor={0.5 0.5 1}, citebordercolor={0.5 1 0.5}, linkcolor=blue]{hyperref}

\begin{document}

\title{Multi-task Deep Learning for Cerebrovascular Disease Classification and MRI-to-PET Translation}

\author{\IEEEauthorblockN{
		Ramy Hussein\IEEEauthorrefmark{1}\IEEEauthorrefmark{5},
		Moss Zhao\IEEEauthorrefmark{1},
		David Shin\IEEEauthorrefmark{2},
		Jia Guo\IEEEauthorrefmark{3},
		Kevin T. Chen\IEEEauthorrefmark{4},
		Rui D. Armindo\IEEEauthorrefmark{1}, 
		Guido Davidzon\IEEEauthorrefmark{1},\\ 
		Michael Moseley\IEEEauthorrefmark{1}, and
		Greg Zaharchuk\IEEEauthorrefmark{1} 		 
	}
	\IEEEauthorblockA{\IEEEauthorrefmark{1}Radiological Sciences Laboratory, Stanford University, Stanford, CA  94305, USA}
	\IEEEauthorblockA{\IEEEauthorrefmark{2}Global MR Applications \& Workflow, GE Healthcare, Menlo Park, CA 94025 , USA}
	\IEEEauthorblockA{\IEEEauthorrefmark{3}Department of Bioengineering, University of California Riverside, CA 92521, USA}	
	\IEEEauthorblockA{\IEEEauthorrefmark{4}Department of Biomedical Engineering, National Taiwan University, Taipei, Taiwan}
	\IEEEauthorblockA{\IEEEauthorrefmark{5}Corresponding author: ramyh@stanford.edu}
}

\maketitle

\begin{abstract}
	Accurate quantification of cerebral blood flow (CBF) is essential for the diagnosis and assessment of cerebrovascular diseases such as Moyamoya, carotid stenosis, aneurysms, and stroke. Positron emission tomography (PET) is currently regarded as the gold standard for the measurement of CBF in the human brain. PET imaging, however, is not widely available because of its prohibitive costs, use of ionizing radiation, and logistical challenges, which require a co-localized cyclotron to deliver the 2 min half-life $^{15}$O radioisotope. Magnetic resonance imaging (MRI), in contrast, is more readily available and does not involve ionizing radiation. In this study, we propose a multi-task learning framework for brain MRI-to-PET translation and disease diagnosis. The proposed framework comprises two prime networks: (1) an attention-based 3D encoder-decoder convolutional neural network  (CNN) that synthesizes high-quality PET CBF maps from multi-contrast MRI images, and (2) a multi-scale 3D CNN that identifies the brain disease corresponding to the input MRI images. Our multi-task framework yields promising results on the task of MRI-to-PET translation, achieving an average structural similarity index (SSIM) of 0.94 and peak signal-to-noise ratio (PSNR) of 38dB on a cohort of 120 subjects. In addition, we show that integrating multiple MRI modalities can improve the clinical diagnosis of brain diseases. 
\end{abstract}


\IEEEpeerreviewmaketitle

\section{Introduction}
\label{intro} 

Cerebrovascular disease refers to conditions that affect the blood vessels and flow of blood in the brain. Cerebrovascular diseases are a leading cause of death or serious long-term disability in the world. In 2019, the Centers for Disease Control and Prevention reported more than 150,000 cerebrovascular-related deaths in the united states \cite{cerebro_stats}, making it the fifth common cause of death. The prompt diagnosis of cerebrovascular diseases is key to faster and more effective treatment to reduce morbidity and mortality.

Cerebral blood flow (CBF) is a measure of the blood supply to a given region of the brain in a given period of time, and has conventional units of ml/100 g/min. Accurate quantification of CBF is essential for the diagnosis and assessment of cerebrovascular disorders (\textit{e.g.}, Moyamoya, ischemic stroke), and neurodegenerative diseases where blood supply to specific regions of the brain is impaired, resulting clinically in different types of dementia \cite{harston2017quantification}. 
Positron emission tomography (PET) with radiolabeled water ($^{15}$O-water, 2 min half-life) is considered the gold standard for measuring CBF in humans \cite{acker2018brain}. PET scans, however, are relatively expensive and not widely available. $^{15}$O-water PET can only be performed in sites where the radioactive substance is produced at a nearby location and can be injected into the bloodstream quickly. Magnetic resonance imaging (MRI), on the other hand, is less expensive, more widely available, and does not involve ionizing radiation. This study aims to improve the clinical utility of MRI-derived CBF measurements, turning brain MRI into PET CBF maps and accurately diagnosing cerebrovascular diseases.

Deep learning models, in particular, the deep convolutional neural networks (CNNs) have opened the door for numerous imaging applications in neuroradiology \cite{zaharchuk2018deep, kaka2021artificial}. For instance, convolutional encoder-decoder architectures have markedly improved the state-of-the-art in neuroimages segmentation \cite{zhang2021deep, abramova2021hemorrhagic, el2021evaluating, dekraker2021surface}, and more importantly, brain image-to-image translation \cite{chen2019ultra, guo2020predicting, armanious2020medgan, yang2020mri}. Moreover, the common deep CNN architectures such as GoogleNet, VGG, ResNet, PyramidNet, and SENet have dramatically boosted the diagnostic and prognostic performance to classify cerebrovascular and neurodegenerative diseases \cite{ho2019machine, dawud2019application}. The unceasing improvement in the performance of emerging neural network architectures will certainly open the door for new commercially available tools for neuroradiologists.

Recently, multi-task neural networks have shown superior performance to other individual neural network architectures on different medical imaging applications \cite{amyar2020multi, von2021multitask}. This type of neural networks simultaneously integrates different pieces of information from diverse tasks to improve the overall performance of the network and leads to better generalization under real-life conditions \cite{ruder2017overview}. In this study, we developed a multi-task CNN architecture for classifying cerebrovascular diseases and synthesizing high-quality PET images from multi-contrast MRI. The proposed joint dual-task model comprises two branches; the first branch adopts a 3D convolutional encoder-decoder network with attention mechanisms to predict the gold standard $^{15}$O-water PET CBF maps from the combination of structural MRI and arterial spin labeling (ASL) MRI perfusion images without using radiotracers. The second branch comprises a multi-scale 3D convolutional network that integrates multi-parametric MRI images to distinguish between healthy controls and people with cerebrovascular diseases. Results show that the proposed multi-task deep learning model can efficiently improve the MRI-to-PET translation performance and the diagnostic accuracy for identifying cerebrovascular disorders.

\section{Related Work}
\label{literature}

In the past three years, several deep convolutional neural networks have been introduced to predict PET CBF maps from structural and perfusion MRI images \cite{guo2020predicting, shin2020ganbert, yousefi2021asl}. In \cite{guo2020predicting}, Guo \textit{et al.} adopted a deep CNN (dCNN) to generate synthetic $^{15}$O-water PET CBF images from multi-parametric MRI inputs including ASL. The dCNN notably improved CBF image quality, when compared to ASL, achieving an average structural similarity index (SSIM) of 0.85. In \cite{shin2020ganbert}, Shin \textit{et al.} studied the possibility of synthesizing different brain PET tracers (specifically AV45, AV1451, fluorodeoxyglucose) solely from T1-weighted MRI images using generative adversarial networks (GANs). This method achieved limited PET prediction results, with an average SSIM of 0.26-0.38. In \cite{yousefi2021asl}, a residual-based attention-guided CNN was introduced for translating 2D ASL and T1-weighted images to PET-like images; achieving an average SSIM of 0.85. Other CNN-based image-to-image translation methods, including encoder-decoder networks and GANs, have been used for MRI-to-computed tomography (CT) translation \cite{kearney2020attention}, CT-to-MRI translation \cite{jin2019deep}, PET-to-CT translation \cite{armanious2019unsupervised}, CT-to-PET translation \cite{ben2019cross}, and also for translating between different MRI neuroimaging modalities \cite{dar2019image}. 

For the detection of cerebrovascular diseases, several deep learning works on MRI and PET neuroimages have emerged in the past few years. In \cite{kim2019machine}, a six-layer convolutional network was proposed to identify Moyamoya disease (MMD) in plain skull radiograph images. The proposed network attained classification accuracy rates of 91\% and 75·9\% for the institutional test set and external validation set, respectively. To improve the diagnostic accuracy of MMD, the authors of \cite{hu2021learning} proposed to use a convolutional-recurrent network architecture that combines a 3D CNN and gated recurrent unit. This hybrid model was able to learn the spatio-temporal features from digital subtraction angiography (DSA) images, showing an average MMD detection accuracy of 97.88\%. Deep learning was also used in \cite{ueda2019deep} to automatically detect cerebral aneurysms from MR angiography images, yielding recognition sensitivity rates between 91-93\%. Further, a deep CNN was adopted for classifying ischemic stroke onset time based on perfusion MRI images \cite{ho2019machine}. The proposed network was able to determine whether the time since stroke (TSS) onset is less than 4.5 hrs with a sensitivity of 0.78 and a negative predictive value of 0.61. In \cite{dawud2019application}, Dawud \textit{et al.} used a pre-trained CNN and transfer learning to identify brain hemorrhage in CT images; revealing high classification accuracy rates of 90.65-93.48\%. 

Recently, multi-task deep learning algorithms have been receiving attention in computer-aided medical applications. In \cite{amyar2020multi}, a multi-task neural network framework was introduced to identify COVID-19 patients and segment abnormal lesions on chest CT images. This model showed impressive results with an AUC score greater than 97\% for the diagnosis task and a Dice score of 0.88 for the segmentation task. Similarly, Von \textit{et al. } in \cite{von2021multitask} developed a multi-task deep learning model for the classification and segmentation of primary bone tumors on musculoskeletal radiographs. This model was able to distinguish between malignant and benign tumors with an average classification accuracy of 80.2\% and segment the bone lesions with an average Dice coefficient of 0.60. Also, multi-task deep learning was adopted for the segmentation and prognosis with head and neck cancer \cite{andrearczyk2021multi}. The proposed multi-task deep UNet model was applied to FDG-PET/CT images to predict patient prognosis and learn the segmentation of head and neck tumors volumes.

\section{Materials and Methods}
\label{method} 

This section describes the dataset and the proposed multi-task network architecture used for simultaneous MRI-to-PET translation and disease classification.

\subsection{Dataset}

This is a retrospective study, approved by the Institutional Review Board of Stanford University in accordance with the ethical standards of the Helsinki declaration for medical research involving human subjects, and HIPAA compliant. Written informed consent was obtained from all participants prior to the study. Data were acquired from 120 subjects (60 healthy controls (HC) and 60 cerebrovascular disease patients) on a 3T PET/MRI hybrid system (SIGNA, GE Healthcare, Waukesha, WI, USA) using an 8-channel head coil. The patients' dataset comprised 52 patients with Moyamoya disease, 4 patients with the intracranial atherosclerotic steno-occlusive disease (ICSD), and 4 patients with stroke. 

The MRI scans included T1-weighted (T1w), T2-weighted fluid-attenuated inversion recovery (T2w-FLAIR), multi-delay pseudo-continuous ASL (PCASL) from which proton density (PD) images are also available, and quantified CBF/arterial transit time (ATT) maps derived from ASL. For all ASL scans, a proton density image and a coil sensitivity map were acquired with a saturation recovery acquisition using TR=2000 ms. For the multi-post labeling delay (PLD) PCASL sequence, crushing gradients (Venc=4cm/s) were adopted to exclude the signal in the arterial component before the 3D spiral readout. All MRI images were co-registered and normalized to the Montreal Neurological Institute (MNI) brain template and resized to 96×96×64 voxels. Quantitative gold standard PET CBF was determined using $^{15}$O-water injection and the image-derived arterial input function kinetic model described in \cite{khalighi2018image}.

Sixty-two participants underwent at least two simultaneous PET and MRI scans, at baseline and 20 minutes after intravenous administration of acetazolamide (a vasodilator that increases the blood flow into the brain). The remaining 58 participants underwent three separate simultaneous PET and MRI scans, two at baseline and one 20 minutes after acetazolamide administration. Acetazolamide was injected during the scan at a dose of 15 mg/kg of body weight with a maximum dose of 1000mg. Eight MRI sequences (T1w, T2w-FLAIR, PD, ATT, single-delay ASL, mean of multi-delay ASL, single-PLD CBF, and multi-PLD CBF) were used as inputs to the model and one $^{15}$O-water PET CBF map served as the ground truth. The total number of scans, before and after acetazolamide administration, is 332. Both input MRI images and output PET images were normalized so that they had a mean intensity of 1 in the whole brain. Data augmentation was also used to enlarge the size of the dataset. The augmentation included flipping, shifting, and rotating the input and output images, resulting in an eight-fold increase in the dataset size.

\subsection{Model Architecture}

\begin{figure*}[tp]
	\centering
	\includegraphics[width=\textwidth]{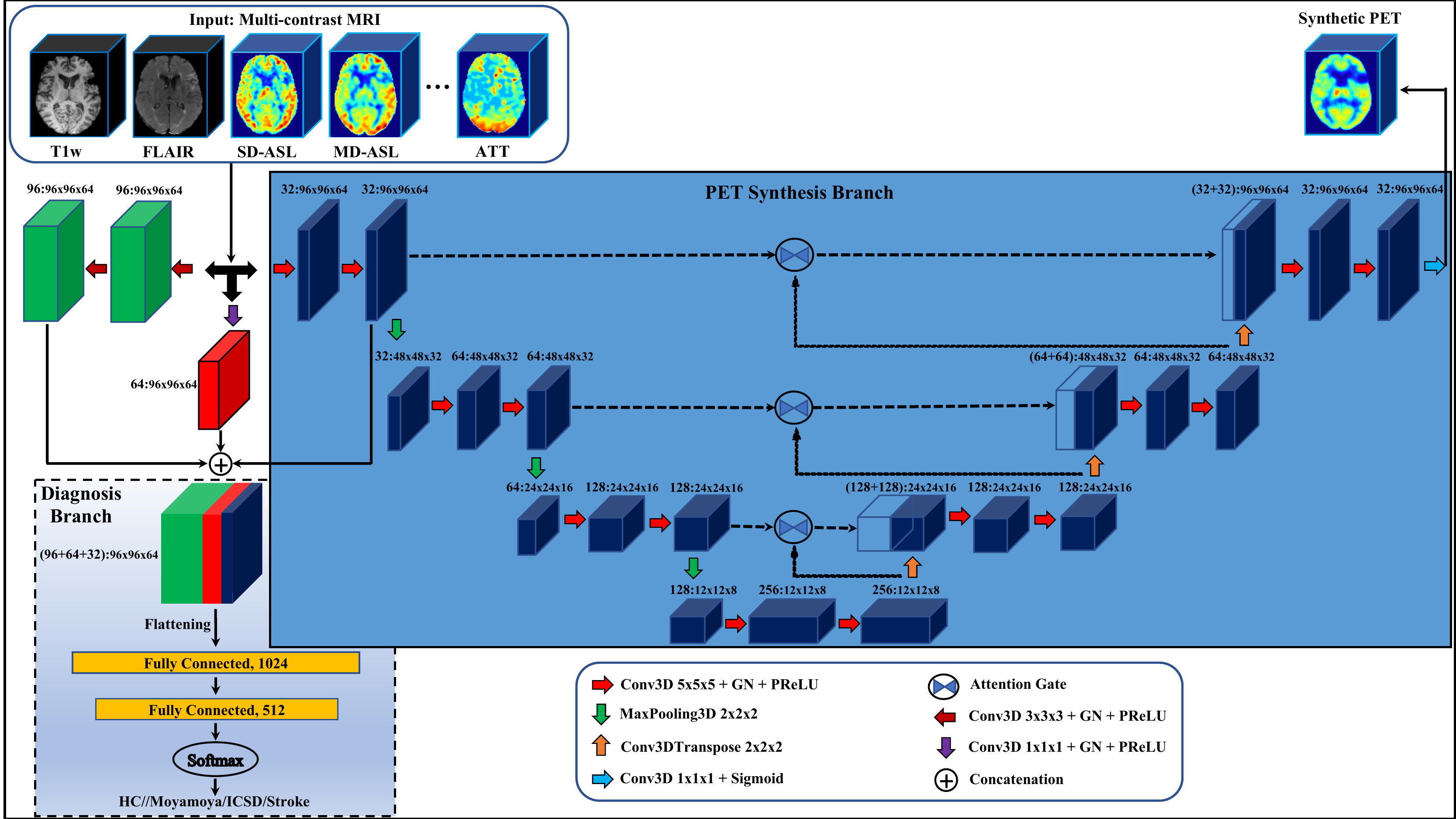}
	\caption{3D multi-task CNN architecture with an attention-based encoder-decoder network for MRI-to-PET translation and a multi-scale convolutional network for cerebrovascular disease classification.}
	\label{fig_model}
\end{figure*}

The proposed 3D multi-task convolutional neural network architecture is depicted in Figure~\ref{fig_model}. The network consists of different branches/sub-networks that use multi-contrast MRI as inputs and are trained simultaneously. In particular, the network incorporates two major branches for improving the clinical utility of MRI-derived CBF measurements: 
(1)~a PET Synthesis Branch is used to transform the structural and ASL perfusion MRI images into PET CBF maps, and
(2)~a Diagnosis Branch is used to classify healthy controls and patients with MMD, ICSD, and stroke.
It is worth highlighting that the MRI-to-PET translation is the prime task of this study, while the classification task is added to ameliorate the extracted feature representations, and hence, improve the quality of synthetic PET images. 

\begin{figure*}[tp]
	\centering
	\includegraphics[width=\textwidth]{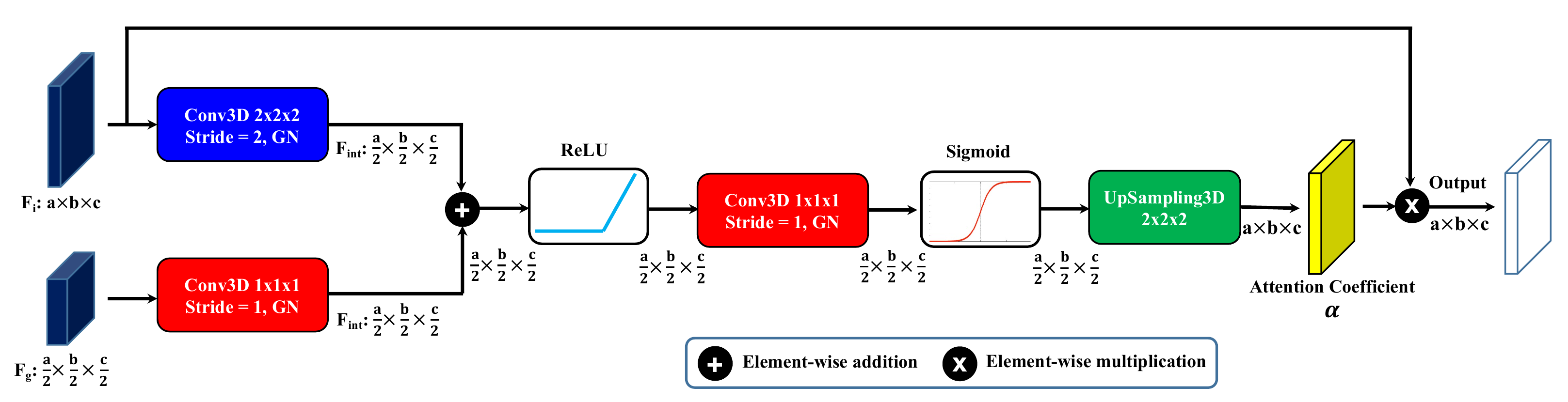}
	\caption{The schematic of attention mechanisms used in the 3D encoder-decoder network.}
	\label{fig_attention}
\end{figure*}

\textbf{PET Synthesis Branch:} A 3D convolutional encoder-decoder network with attention mechanisms was adopted to integrate spatial data across multiple MRI image types for synthetizing PET CBF maps, as shown in Figure 1. Both the encoder and decoder modules use 3D CNNs, where the encoder compresses the input MRI images into a more condensed representation, and the decoder uses this representation to output PET-like images. Since different MRI sequences and spatial patterns impose different effects on the quality of synthesized PET images, the attention mechanism, shown in Figure\ref{fig_attention}, is embedded into the encoder-decoder network to concurrently search the relevant aspects of the input at the channel and spatial levels for a fine-grained quality prediction. 
The encoder-decoder network is trained with a customized loss function, computed as:
\begin{equation}\label{Eqn_translation_loss}
	\small
	L_{trans} = w_1  MSE + w_2  MAE  + w_3 (1-SSIM) + w_4 PSNR
\end{equation}
where $ L_{trans} $ is the MRI-to-PET Translation loss; $ w_1$,$ w_2 $, $ w_3 $, and $ w_4 $ are weights that take values between 0 and 1; $ MSE $,  $ MAE $, $ SSIM $, $ PSNR $ refer to the mean squared error, mean absolute error, structural similarity index, and peak signal-to-noise ratio, respectively, and they are defined as:
\begin{equation}
	\small
	MSE  = \dfrac{1}{mnp} \sum\limits_{i=1}^{m} \sum\limits_{j=1}^{n} \sum\limits_{k=1}^{p}  \Big( x(i,j,k)- y( i,j,k)   \Big) ^2
\end{equation}	
where $ x $ and $ y $ refer to the \textit{true} and \textit{predicted} PET images, $ m $, $ n $, and $ p $ are the dimensions of the 3D PET images.		
\begin{equation}
	\small
	MAE  =  \dfrac{1}{mnp} \sum\limits_{i=1}^{m} \sum\limits_{j=1}^{n} \sum\limits_{k=1}^{p}  \Big| x(i,j,k)- y( i,j,k)   \Big|
\end{equation}
\begin{equation}
	\small
	SSIM  =  \dfrac{  \left( 2 \mu_x \mu_y + c_1 \right) \left(  2 \sigma_{xy} + c_2 \right) }{    \left( \mu_x^2 + \mu_y^2  + c_1\right) \left(  \sigma_x^2 + \sigma_y^2  + c_2 \right)  } 
\end{equation}
where $ \mu_x $ is the mean of $ x $, $ \mu_y $ is the mean of $ y $, $  \sigma_x^2 $ is the variance of $ x $, $  \sigma_y^2 $ is the variance of $ y $,  $ \sigma_{xy} $ is the covariance of $ x $ and $ y $, $c_1$=$(0.01 c_{max})^2$ and $c_2$=$(0.03 c_{max})^2$ are two constants to stabilize the division with weak denominator, and $c_{max}$ denotes the maximum intensity value of the image. 
\begin{equation}
	\small
	PSNR = 10 \cdot \log_{10} \Big( \dfrac{c_{max}}{MSE} \Big)
\end{equation}

\textbf{Diagnosis Branch:} A multi-scale 3D convolutional neural network is adopted to distinguish between healthy controls and patients with MMD, ICSD, and stroke. The multi-parametric MRI images are fed into three parallel paths of 3D convolution layers with different kernel sizes, which allows learning local features (through smaller convolutions) and high-abstracted features (with larger convolutions) (see Figure~\ref{fig_model}). The extracted multi-scale feature representations are then concatenated into a single feature tensor forming the input of the next few layers. The aggregated output is flattened and presented as an input to two fully connected layers, and then the softmax function is used to compute the label probabilities and prediction. The categorical cross-entropy function is used as a loss function for the cerebrovascular disease classification task, which is defined as:
\begin{equation}\label{Eqn_classification_loss}
	L_{class} = - \dfrac{1}{N} \sum\limits_{i=1}^{N} \sum\limits_{c=1}^{C} \mathds{1}_{y_i \in C_c} \log_{10} \big( P_{model} \left[ y_i \in C_c \right]  \big)
\end{equation}
where $L_{class}$ is the classification loss, $N$ is the number of observations, $ C $ is the number of classes, $ \mathds{1} $ is an indicator function (0 or 1) of the observation $ i $ belonging to the class $ c $, and $ P_{model} \left[ y_i \in C_c \right] $ is the probability that observation $ i $ belongs to class $ c $. 

From equations \ref{Eqn_translation_loss} and \ref{Eqn_classification_loss}, the global loss function ($ L_{global} $) for both MRI-to-PET translation and disease classification tasks is computed as:
\begin{equation}
	L_{global} =  L_{trans} +  L_{class}
\end{equation}

In our experiments, the Nesterov Adam optimizer \cite{dozat2016incorporating}, an improved variant of the Adam optimization algorithm, was used with a learning rate of 0.0002 and a batch size of 4. The proposed multi-task neural network was trained with the global loss function for 150 epochs and early stopping of 20 epochs.

The proposed multi-task network was trained and tested using fourfold cross-validation. The dataset was divided into four subgroups, each includes PET and MRI images from 15 healthy control participants (with 40 scans), 13 patients with Moyamoya disease (with 38 scans), one ICSD patient (with 3 scans), and one stroke patient (with 2 scans). For each fold, the scans from three of the four sub-groups were used for training, from which 10\% were randomly selected for validation. The fourth subgroup was then used for testing the performance of the trained multi-task network. To avoid data leakage, we were careful to avoid having a single subject’s scans (either baseline or post-acetazolamide) simultaneously in the training and test sets.

\section{Results and Discussion}
\label{results}

The performance of the proposed multi-task network was quantitatively evaluated using SSIM, normalized root-mean-square error (NRMSE), and PSNR for the MRI-to-PET Translation task. The performance metrics of accuracy (Acc), sensitivity (Sens), specificity (Spec), precision (Prec), false-positive rate (FPR), false-negative rate (FNR), and Matthew's correlation coefficient (MCC) were also used for the disease classification task.

\begin{figure*}[tp]
	\centering
	\includegraphics[width=\textwidth]{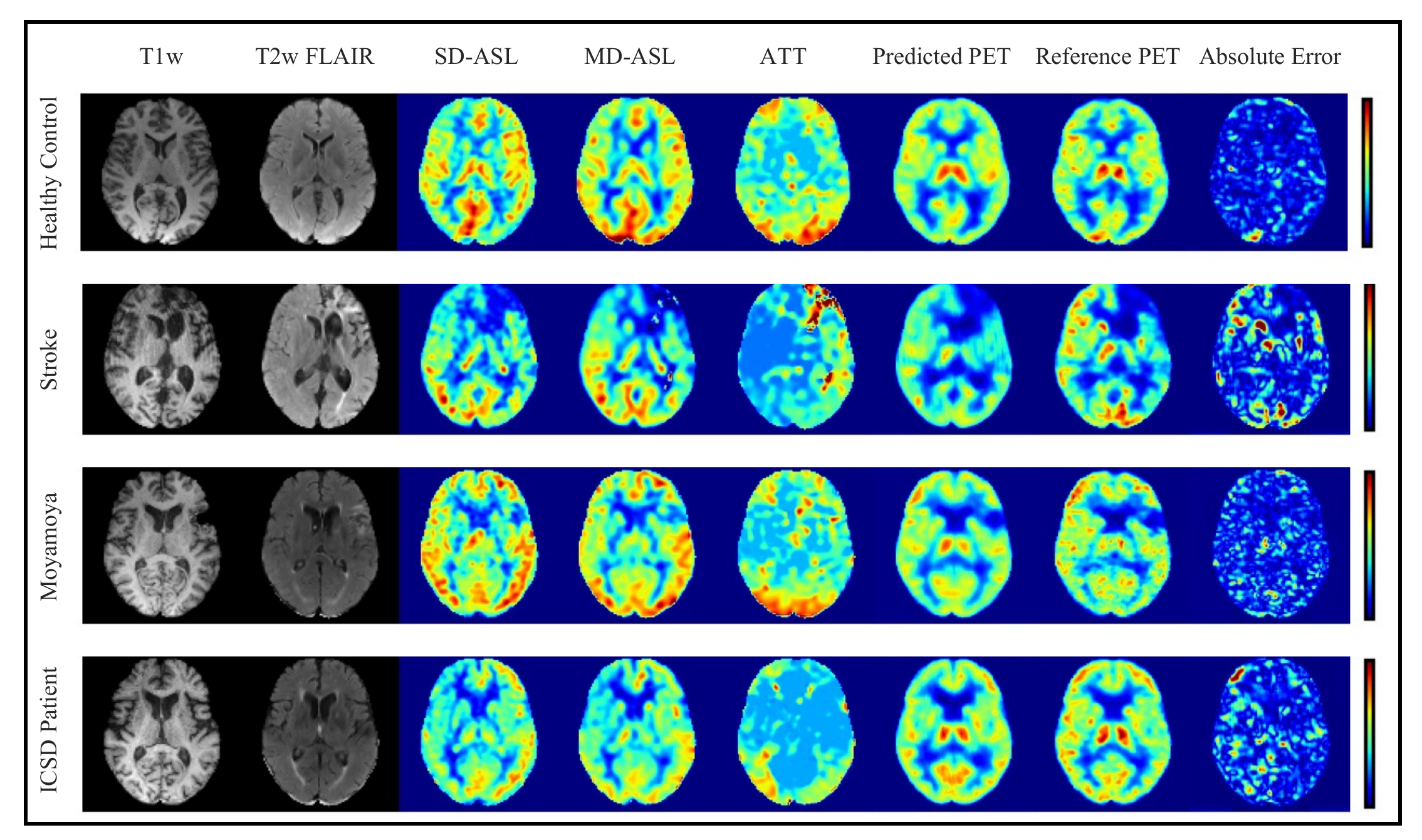}
	\caption{MRI-to-PET prediction results for healthy control and cerebrovascular disease patients: Examples of input MRI sequences, output synthetic PET, target PET, and corresponding Absolute Error maps.}
	\label{fig_prediction_axial}
\end{figure*}

\textbf{MRI-to-PET Translation Results:} Compared to a previous PET CBF prediction method that achieved an average SSIM of 0.85 and NRMSE of 0.209 \cite{guo2020predicting}, our model yielded improved PET prediction performance, achieving an average SSIM of 0.94, NRMSE of 0.038, and also PSNR of 38.8dB. Figure~\ref{fig_prediction_axial} shows the input MRI volumes, predicted (synthetized) PET CBF maps, gold standard PET CBF measurements, and corresponding absolute error maps (magnified) for healthy controls and patients with cerebrovascular diseases. Results indicate that a 3D convolutional encoder-decoder network integrating multi-contrast information from brain structural MRI and ASL perfusion images can efficiently synthesize high-quality PET CBF maps without the need for radiotracers. It also shows how well-designed loss functions and attention mechanisms can improve the PET CBF prediction results. Using grid search, the weights ($w_1$, $w_2$, $w_3$, and $w_4$) of $L_{trans}$ (the loss function of the MRI-to-PET translation task) were assigned to 0.15, 0.15, 0.60, and 0.20, respectively. The highest weight went to the SSIM loss in view of the fact that SSIM measures the perceptual difference between two images.

\begin{figure}[tp]
	\centering
	\includegraphics[width=\columnwidth]{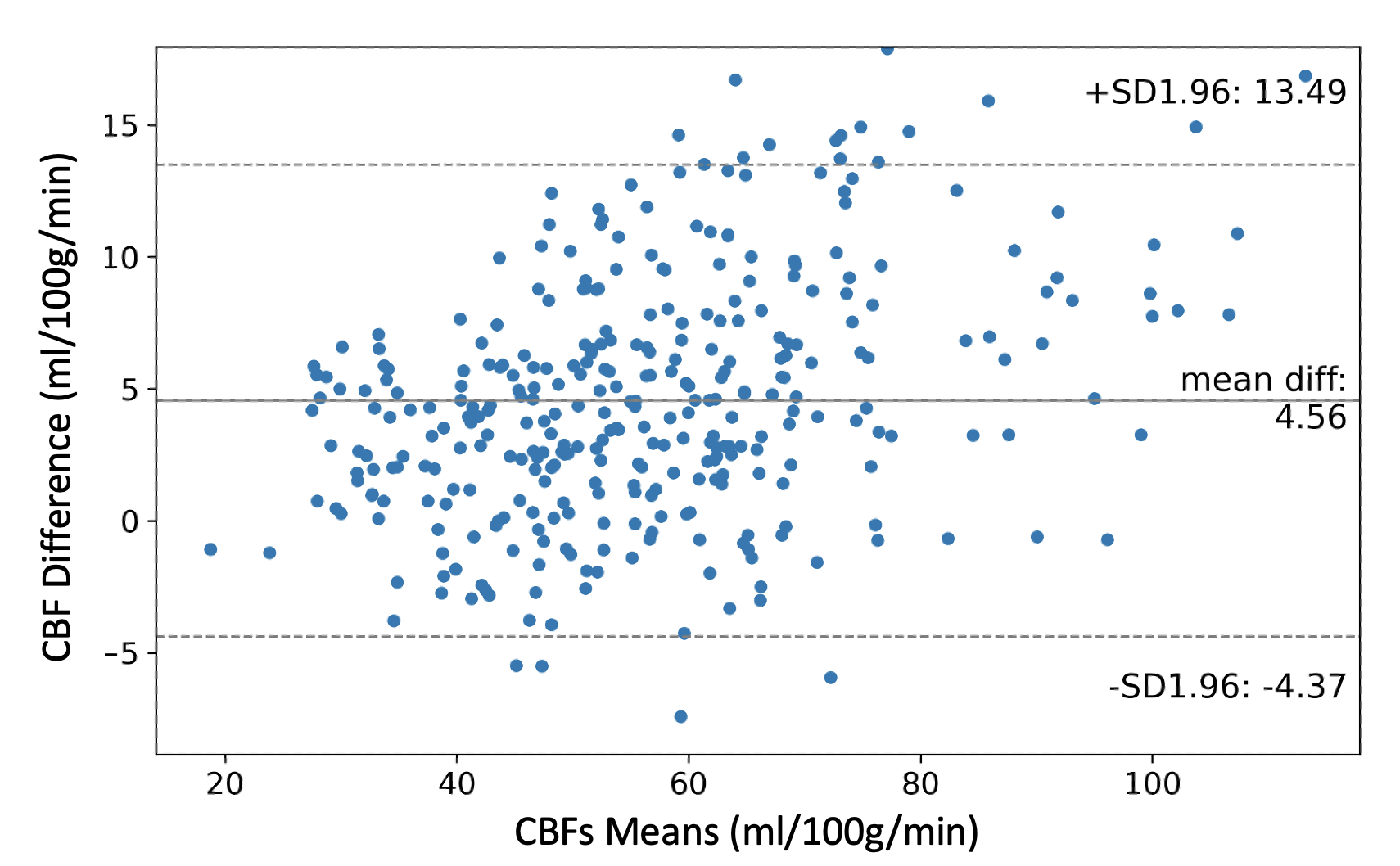}
	\caption{Bland-Altman plot of whole brain CBF Mean and Difference of measured and synthetic PET.}
	\label{fig_blandaltman}
\end{figure}

\begin{figure}[tp]
	\centering
	\includegraphics[width=0.98\columnwidth, height=0.8\columnwidth]{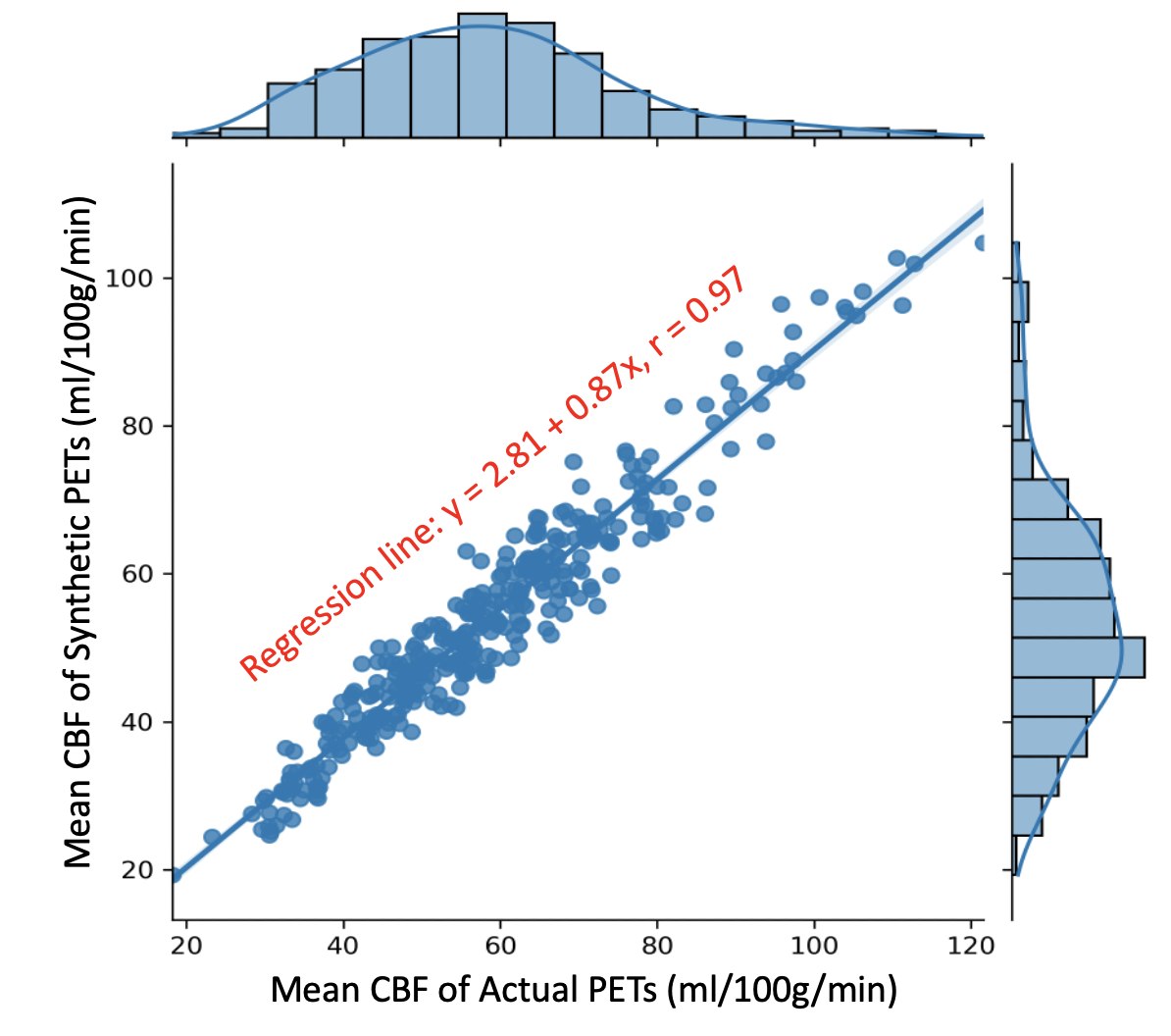}
	\caption{Joint plot of mean global CBF in measured and synthetic PET.}
	\label{fig_jointplot}
\end{figure}

To assess the clinical significance of the proposed PET prediction algorithm, a set of paired comparison analyses were conducted. The Bland-Altman plot in Figure~\ref{fig_blandaltman} delineates the agreement between the mean CBF of the true and predicted PET CBF maps. It shows a small bias, where the true gold standard PET CBF measurements in the whole brain are 4.6 ml/100g/min higher than the synthetic PET CBF maps produced by our encoder-decoder network, with 95\% confidence intervals of -4.4 and +13.5 ml/100g/min. Figure~\ref{fig_jointplot} describes the histogram, density plots and joint plot of the mean CBFs of true and synthetic PET images, showing high levels of agreement and correlation (Pearson's correlation coefficient = 0.97). 

\textbf{Cerebrovascular Disease Classification Results:} This section depicts how 3D multi-scale convolutional networks, as part of the proposed multi-task deep learning model, can differentiate between healthy controls and patients with Moyamoya disease, intracranial steno-occlusive disease, and stroke. 

\begin{figure}[tp]
	\centering
	\includegraphics[width=0.95\columnwidth]{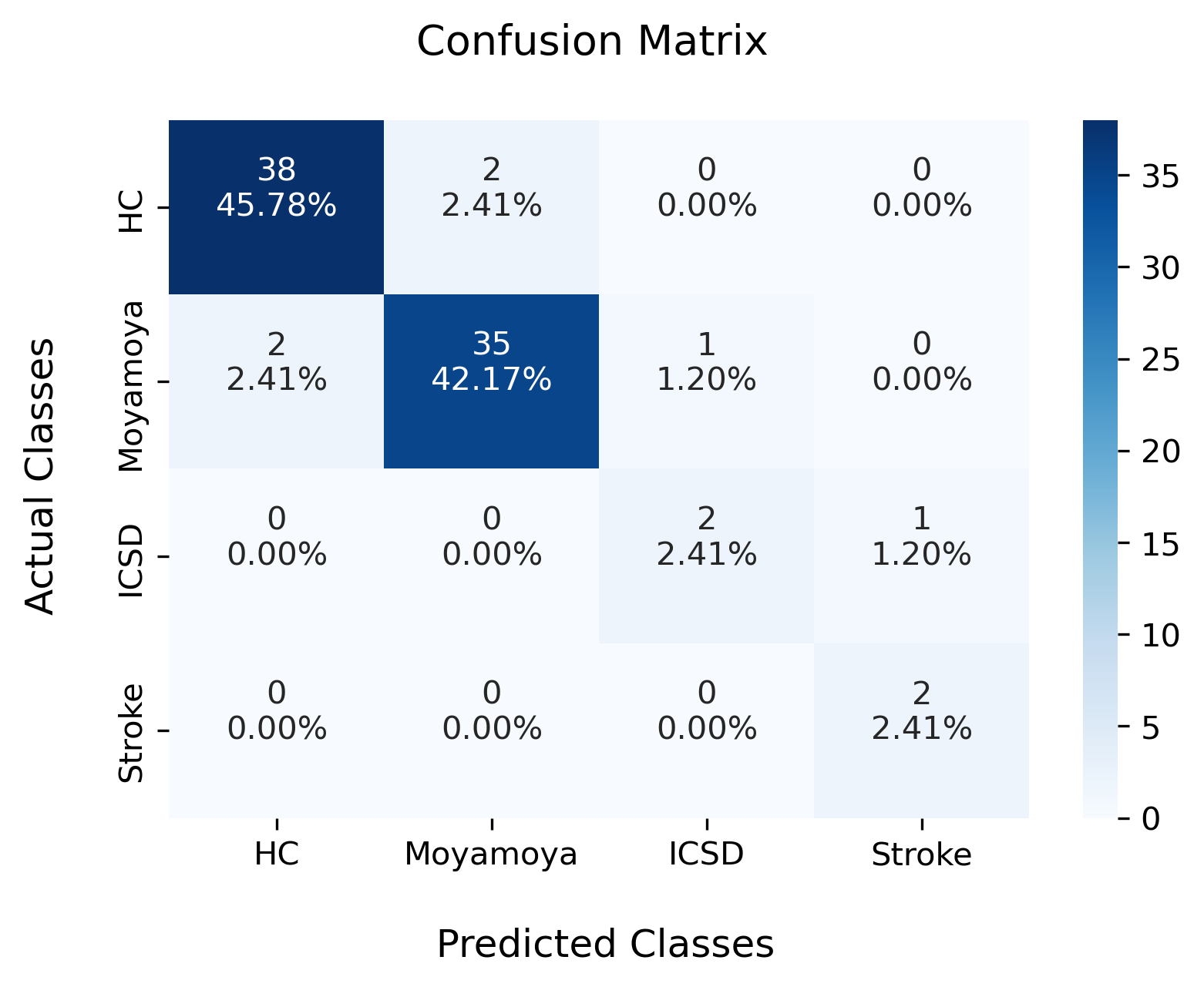}
	\caption{Confusion matrix of the multi-task convolutional nertwork.}
	\label{fig_conf_matrix}
\end{figure}

Figure~\ref{fig_conf_matrix} depicts the confusion matrix of the proposed multi-task network for one of the four sub-groups of data. One can observe that 38 of 40 healthy control scans are classified correctly, where only 2 scans are misclassified as Moyamoya patients (both scans were for the same participant). For Moyamoya disease, only two scans are misidentified as healthy controls and one misidentified as an intracranial steno-occlusive disease. The remaining 35 scans are identified correctly. The classification of ICSD and stroke was more challenging because of the limited number of participants having these diseases. We used data augmentation to generate variants of the multi-contrast MRI images and corresponding PET CBF measurements, which helped expand the training dataset and improve the classification performance of the model and its ability to generalize in clinical settings. 

The test set included one ICSD patient with three scans (two scans at the baseline and one scan after acetazolamide administration). The trained network was able to identify the two baseline scans correctly where the post-acetazolamide scan was identified as Stroke. It is worth clarifying that ICSD occurs when blood flow to the brain is restricted by narrowed arteries or plaque buildup, and without adequate treatment, ICSD patients may develop mini-strokes or strokes \cite{choi2007detection}. This may explain why a post-acetazolamide scan for an ICSD patient could be classified as a stroke. Lastly, the model was tested on a stroke patient with 2 scans (one pre-acetazolamide and one post-acetazolamide), and both scans were classified correctly. 

Table~\ref{table_diagnosis_results} reports the performance metrics of the multi-task network for the classification of healthy controls and patients with Moyamoya disease, ICSD, and stroke. An average classification accuracy, sensitivity, and specificity of 96.38\%, 88.44\%, and 97.10\% were achieved, respectively. The FPR, FNR, and MMC were also evaluated for this imbalanced classification problem, revealing a notably high performance of 0.028, 0.115, and 0.812, respectively. This gives us insights into how multi-task deep learning and sharing feature representations among two tasks can help maintain reliable performance even with limited data.

\begin{table}[!t]
	\caption{Performance of the multi-task DL model for identification of healthy controls and patients with Moyamoya disease, ICSD and Stroke.}
	\label{table_diagnosis_results}
	\centering
	\small
	\begin{tabular}{|c||c|c|c|c||c|}
		\hline
		\begin{tabular}[c]{@{}c@{}}Classification\\ metrics\end{tabular} & HC & MMD & ICSD  & Stroke & Average \\
		\hline
		Acc (\%)                                          & 95.18           & 93.97    & 97.60 & 98.80  & 96.38   \\
		\hline
		Sens (\%)                                                      & 95.00           & 92.10    & 66.66 & 100.0  & 88.44   \\
		\hline
		Spec (\%)                                                       & 95.34           & 95.55    & 98.75 & 98.76  & 97.10   \\
		\hline
		Prec (\%)                                                          & 95.00           & 94.59    & 66.66 & 66.66  & 80.73   \\
		\hline
		FPR                                                              & 0.046           & 0.044    & 0.012 & 0.012  & 0.028 \\
		\hline
		FNR                                                              & 0.050           & 0.079    & 0.333 & 0.0    & 0.115   \\
		\hline
		MCC                                                              & 0.903           & 0.878    & 0.654 & 0.811  & 0.812 \\
		\hline 
	\end{tabular}
\end{table}

Although the proposed model produces high-quality synthetic PET images, the neuroradiologists still need to spend substantial time examining 3D images trying to identify the area(s) of the brain affected by the cerebrovascular disease. In future work, we plan to add a third branch to automatically localize the brain regions with abnormally low cerebral blood flow. An expansion of this approach to other types of brain diseases (like certain types of dementia) could also be clinically valuable. 

\section{Conclusion}

Adequate quantification of PET from MRI has a great potential for increasing the accessibility of cerebrovascular diseases assessment for underserved populations and underprivileged communities. In this study, we proposed a multi-task deep learning model that allows for accurate and simultaneous brain MRI-to-PET translation and classification of cerebrovascular diseases. The network consists of two branches that cooperatively use structural MRI and ASL perfusion images as inputs. The first branch adopted an attention-guided 3D convolutional encoder-decoder network that efficiently synthesizes high-quality PET CBF maps from multi-contrast MRI while eliminating the need for radioactive tracers. The second branch used a multi-scale convolutional neural network to extract the distinguishable imaging biomarkers and thus differentiate between healthy controls and patients with cerebrovascular diseases. The proposed multi-task learning approach was found to achieve superior performance than existing medical image-to-image translation and classification techniques.

\medskip
{
	\small
	\bibliographystyle{IEEEtran}
	\bibliography{refs}
}

\end{document}